\documentclass{article}

\PassOptionsToPackage{numbers, compress}{natbib}

\usepackage[preprint]{style}

\usepackage[utf8]{inputenc}
\usepackage[T1]{fontenc}  
\usepackage{hyperref}     
\hypersetup{
  colorlinks   = true, 
  urlcolor     = blue,
  linkcolor    = blue,
  citecolor   = blue
}
\usepackage{url}         
\usepackage{booktabs}     
\usepackage{amsfonts}   
\usepackage{amsmath,amssymb,amsthm}       
\usepackage{nicefrac}    
\usepackage{microtype}    
\usepackage{xcolor}        
\usepackage{multirow}
\usepackage{braket}
\usepackage{graphicx}

\title{Hybrid Parameterized Quantum States for \\ Variational Quantum Learning}

\author{%
  Chen-Yu Liu \\
  National Taiwan University \\
  \\
}

\begin{document}

\maketitle

\begin{abstract}

Variational quantum learning faces practical challenges in the noisy intermediate-scale quantum (NISQ) era. Parameterized quantum circuit (PQC) models suffer from statistical uncertainty due to finite-shot measurements and are highly sensitive to quantum noise, while purely classical approximations like neural quantum states (NQS) lack access to genuine quantum correlations and are limited in scalability. This work introduces Hybrid Parameterized Quantum States (HPQS), a general-purpose modeling framework that interpolates between quantum and classical parameterizations. HPQS combines PQC-based measurements with neural estimators via a blending mechanism and postprocessing functions, enabling enhanced, shot-efficient evaluation under hardware constraints. We demonstrate HPQS across three representative quantum learning tasks: (1) Expectation-based QML, where HPQS yields higher classification accuracy than PQC-only and NQS-only baselines under limited quantum measurements. (2) Quantum-Train, where HPQS generates the entire parameter set of classical networks using polylogarithmic trainable variables; and (3) Quantum Parameter Adaptation (QPA), where HPQS produces LoRA adapter parameters for fine-tuning large language models like GPT-2 and Gemma-2 with improved perplexity under low-shot conditions; Together, these results position HPQS as a scalable, noise-resilient approach for variational quantum learning, compatible with both current NISQ hardware and future fault-tolerant architectures.

\end{abstract}

\section{Introduction}

Quantum machine learning (QML) is a rapidly growing field that seeks to leverage the representational power of quantum systems to improve learning performance in classification, generative modeling, and reinforcement learning tasks \cite{perez2020data, schuld2021effect, schuld2019quantum, liu2021hybrid, biamonte2017quantum}. Among various approaches, \textit{variational quantum algorithms (VQAs)} have emerged as a central paradigm due to their compatibility with near-term quantum hardware. Despite inherent limitations in circuit width and depth, they have demonstrated potential quantum advantages on carefully constructed problem instances \cite{riste2017demonstration, huang2022quantum, cerezo2022challenges}. These methods rely on \textit{parameterized quantum circuits (PQCs)} to represent learnable quantum states, where model training is carried out by adjusting circuit parameters to minimize a loss function through repeated quantum measurements. 
However, the practical deployment of PQC-based models faces two critical limitations:  
(1) {the inefficiency of finite-shot measurements}, which introduces statistical noise and limits the accuracies of probability and expectation value estimation \cite{cerezo2022challenges, banchi2024few}, and  
(2) {the scalability bottleneck of circuit size and depth}, constrained by hardware fidelity and decoherence in today’s noisy intermediate-scale quantum (NISQ) systems \cite{preskill2018quantum}.

\begin{figure}[h]
\centering
\includegraphics[width=\linewidth]{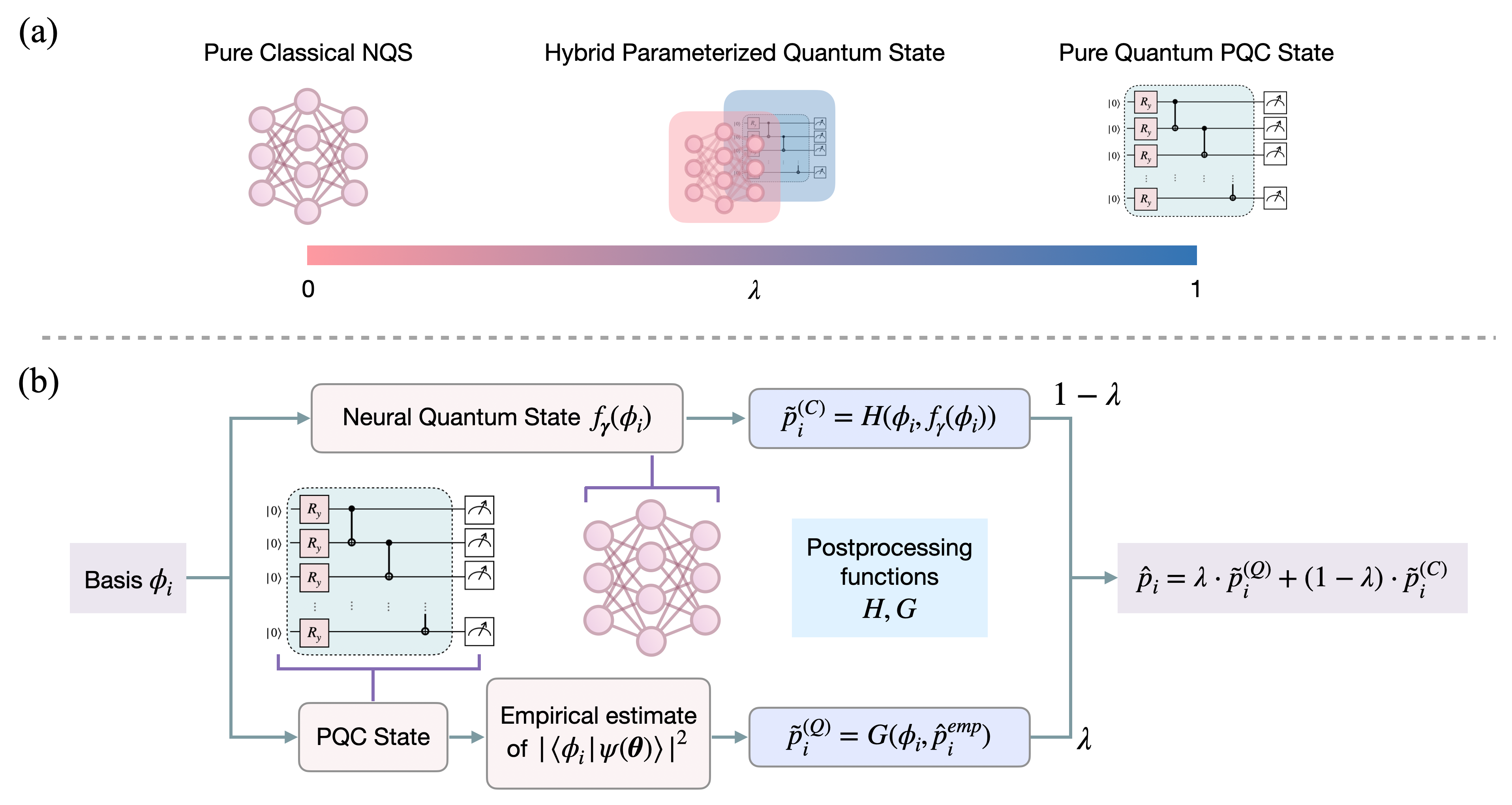}
\caption{Overview of the HPQS framework.
(a) HPQS interpolates between a pure classical NQS and a pure quantum PQC state using a blending coefficient $\lambda \in [0,1]$.
(b) Architecture of HPQS: Given a basis state $\phi_i$, both a classical NQS $f_{\boldsymbol{\gamma}}(\phi_i)$ and a PQC-based quantum state $|\psi(\boldsymbol{\theta})\rangle$ are evaluated. The classical and quantum outputs are then postprocessed by functions $H$ and $G$, respectively, to produce $\tilde{p}_i^{(C)}$ and $\tilde{p}_i^{(Q)}$. The final probability estimate $\hat{p}_i$ is computed as a weighted combination of the two, enabling robust and shot-efficient variational quantum learning.}
 \label{fig:scheme}
\end{figure}

Recent efforts have proposed solutions such as classical shadow tomography for more efficient expectation estimation \cite{aaronson2018shadow, christandl2012reliable}, error mitigation methods to reduce noise bias \cite{cai2023quantum, endo2018practical}, and quantum error correction (QEC) techniques that protect logical qubits from noise \cite{lidar2013quantum, bravyi2024high}. While these approaches show promise, they often introduce additional layers of computational or hardware complexity. For instance, QEC schemes typically require tens to hundreds of physical qubits to encode a single logical qubit, making them challenging to deploy in practical large-scale learning settings with current hardware.

While QML has predominantly focused on quantum circuits, in the classical computational perspective of quantum system simulation, \textit{Neural Quantum States (NQS)}, which are classical neural network models that approximate quantum state distributions over computational basis states, has been proposed \cite{carleo2017solving, glasser2018neural, hartmann2019neural}. NQS-based models are powerful, expressive, and easy to optimize using backpropagation. They have shown promising results in simulating quantum systems and generative modeling \cite{choo2018symmetries, torlai2018neural}.  Given the nature of pure classical simulation, NQS lacks access to real quantum data and offers no inherent guarantee that the modeled distributions reflect physical quantum states unless specifically constrained \cite{choo2020fermionic, kim2024neural, stokes2020phases}.

From the perspective of PQC-based models, incorporating the inductive structure of NQS can alleviate the statistical inefficiency introduced by finite-shot quantum measurements. Classical neural networks can be trained to concurrently estimate measurement probabilities, providing a smooth, differentiable, and noise-free approximation of the quantum state. This not only {potentially} improves learning efficiency under limited quantum samples but also mitigates the impact of quantum hardware noise during training.
Conversely, from the perspective of NQS-based models, introducing PQC components allows the network to leverage physically grounded quantum information that naturally captures entanglement, superposition, and other structural correlations inherent in quantum systems. The representation induced by PQC is known to be highly expressive with a relatively small number of parameters \cite{chen2020variational, du2020expressive, benedetti2019parameterized}.
As we demonstrate in this work, the complementary interaction between classical and quantum components leads to empirical performance gains when either neural or quantum parameters are incorporated.

Nevertheless, in the low-qubit regime (dozens of qubits) that is possible to be approximated by NQS efficiently and under {NISQ} condition of quantum system, it becomes promising to \textbf{combine finite-measurements PQC with NQS}. This enables hybrid models that both learn from classical approximations and incorporate partial quantum state information. Such hybridization is particularly compelling when only a subset of measurement outcomes is available or when the number of shots is limited.

In this work, we introduce \textit{Hybrid Parameterized Quantum States (HPQS)}, a variational modeling framework that unifies the expressiveness of NQS with the physical grounding of PQC states. HPQS defines a learnable quantum-classical state that integrates empirical quantum measurements with classical neural estimators, enabling noise-resilient and shot-efficient variational learning under sparse, noisy, or limited-access quantum observations. Key contributions of this work are as follows:
\begin{itemize}
    \item \textbf{Hybrid Formulation}: HPQS is formulated as a new class of variational quantum states that interpolates between PQC and NQS, enabling enhanced estimation of measurement probabilities using partial quantum measurements and a classical neural estimator.
    
    \item \textbf{Generality}: The HPQS framework generalizes both PQC-only and NQS-only models as special cases and can be further reduced to other hybrid quantum-classical frameworks under certain conditions (Appendix~\ref{app:otherquantum}), demonstrating that HPQS serves as a unifying and generalizable framework. 

    \item \textbf{Demonstrated Effectiveness}: 

    The effectiveness of HPQS is demonstrated across three representative QML scenarios: two recently proposed tasks related to Quantum-Train, and one standard benchmark task involving expectation-based quantum classification.
    \begin{itemize}
        \item In \textbf{Quantum-Train (QT)}, HPQS generates the full parameter set of classical neural networks using only polylogarithmic trainable variables.
        \item In \textbf{Expectation-Based QML}, HPQS enables more stable and enhanced expectation value estimation under low-shot measurement regimes.
        \item In \textbf{Quantum Parameter Adaptation (QPA)}, HPQS surpasses pure NQS and PQC settings with finite measurement shots in generating LoRA adapter weights for large language model fine-tuning task.

    \end{itemize}
    \item \textbf{Long-Term Viability}: HPQS is positioned as a practical yet forward-compatible architecture for quantum–classical co-design, suitable for both current NISQ-era systems and future fault-tolerant quantum hardware. Notably, even fault-tolerant devices are constrained by finite-shot measurements, and HPQS remains beneficial in such regimes by enhancing estimation efficiency.
\end{itemize}

\section{Related works}

\subsection{Parameterized Quantum Circuit States}

PQC states form the cornerstone of variational quantum learning \cite{chen2020variational, du2020expressive, benedetti2019parameterized}. In this framework, a quantum state is defined by a parameterized ansatz, typically composed of alternating layers of entangling gates (e.g., CNOT) and single-qubit rotations (e.g., $R_Y$, $R_Z$). For $n$ qubits, a PQC defines a variational quantum state of the form:
\begin{equation}
|{\psi(\boldsymbol{\theta})} \rangle = U(\boldsymbol{\theta}) |{0}
\rangle^{\otimes n},
\end{equation}
where $U(\boldsymbol{\theta})$ is a trainable quantum circuit parameterized by a vector of real-valued parameters $\boldsymbol{\theta}$ and $|0\rangle^{\otimes n}$ is an $n$-qubit initialized state. The optimization objective typically minimizes a classical loss function $\mathcal{L}$ that depends on either measurement probabilities or expectation values of observables.
PQC-based quantum states offer strong expressivity and can be designed to be hardware-efficient, particularly when adapted to the constraints of NISQ devices \cite{kandala2017hardware, leone2024practical}.
However, their practical deployment faces two major bottlenecks: (1) the \textit{finite-shot measurement problem}, which introduces statistical uncertainty in estimating measurement probabilities or expectation values; and (2) the presence of hardware-induced quantum noise, which further distorts the observed distributions.

In an exact classical simulation of a variational quantum state \( |\psi(\boldsymbol{\theta})\rangle \), the probability of measuring a computational basis state \( |\phi_i\rangle \) is given precisely by calculating \( |\langle \phi_i | \psi(\boldsymbol{\theta}) \rangle|^2 \). However, on real quantum hardware, this probability must be estimated via repeated sampling (shots), which introduces statistical fluctuations. Let \( \hat{P}(|\phi_i\rangle) \) denote the empirical estimate of this probability after \( n'_{\text{shot}} \) measurement repetitions.
The expression \( n'_{\text{shot}} = n_{\text{shot}} / 2^n \) provides an approximate scale for the average number of shots per basis state, where \( n_{\text{shot}} \) is the total number of measurements and \( 2^n \) is the Hilbert space dimension. This estimate is intended only as a guideline for scaling behavior, since measurement outcomes are not uniformly distributed and many basis states may receive few or no samples in practice.
Hoeffding’s inequality provides an upper bound on the probability that this estimate deviates from the true value by at least \( \epsilon > 0 \):
\begin{equation}
P\left( \left| \hat{P}(|\phi_i\rangle) - \mathbb{E}[P(|\phi_i\rangle)] \right| \geq \epsilon \right) \leq 2 \exp(-2 \epsilon^2 n'_{\text{shot}}).
\end{equation}
This bound highlights a key limitation of PQC-based methods: accurately estimating the measurement probability of each basis state requires a large number of shots, particularly when the desired precision \( \epsilon \) is small or the number of relevant basis states grows exponentially with the number of qubits (see Appendix~\ref{app:err}). This statistical bottleneck has motivated the development of shot-efficient quantum state estimation techniques, such as shadow tomography \cite{christandl2012reliable, cramer2010efficient, hsu2024quantum}, which aim to infer properties of quantum states from a limited number of measurements. While these methods can reconstruct global information from fewer samples, their primary application has focused on estimating the expectation values of observables rather than recovering the full probability distribution over basis states.

Another major challenge is the presence of \textit{quantum noise}, which arises from gate imperfections, decoherence, and readout errors in NISQ-era devices \cite{preskill2018quantum, kandala2019error}. As quantum circuits grow in depth or width, the accumulated noise compounds, leading to degraded output fidelity and unreliable measurements. This effect is particularly problematic in variational algorithms, where the learning signal (e.g., gradients or expectation values) may be obscured or biased by stochastic hardware noise. In practice, mitigating these errors often requires significant overhead, such as repetition for averaging, error correction codes, or post-processing techniques, all of which increase computational and physical resource demands \cite{cai2023quantum, endo2018practical, lidar2013quantum, bravyi2024high}.

Despite these limitations, PQC states remain a powerful tool in quantum machine learning. They naturally encode quantum correlations such as entanglement and superposition, which are difficult to capture using classical neural networks alone. 
\subsection{Neural Quantum States}
NQS' are classical, parameterized models designed to approximate quantum probability distributions. Introduced in the context of variational Monte Carlo for quantum many-body systems \cite{carleo2017solving, glasser2018neural, hartmann2019neural, choo2020fermionic, kim2024neural, stokes2020phases}, NQS represent the wavefunction amplitudes or measurement probabilities (we use measurement probabilities in this work) using a classical neural network $f_{\gamma}(\phi_i)$, where $\phi_i$ denotes the basis state (typically expressed in binary vector) and $\gamma$ are the learnable parameters of the network.
In the most common formulation, the NQS defines a probability distribution over the computational basis as:
\begin{equation}
p_{\text{NQS}}(\phi_i) = \frac{f_{\gamma}(\phi_i)}{\sum_j f_{\gamma}(\phi_j)},
\end{equation}
where $f_{\gamma}(\cdot)$ is a positive-valued neural network (e.g., using softplus or sigmoid activations). The network is trained to approximate the target quantum distribution by minimizing a loss function derived from observed data. 

NQS models offer several practical advantages. They are fully differentiable and can be optimized using standard gradient-based techniques such as stochastic gradient descent (SGD). Implemented on classical hardware, an NQS defines a function $f_{\gamma} : \{0,1\}^n \rightarrow \mathbb{R}_{\geq 0}$, where $\gamma$ denotes the set of neural network parameters and $n$ is the number of qubits (or bits in the basis state $\phi_i$).
Thanks to the universal approximation theorem \citep{cybenko1989approximation, hornik1989multilayer}, a sufficiently expressive NQS (e.g., a multilayer perceptron with non-linear activations) can approximate any target distribution $p^*(\phi_i)$ to arbitrary accuracy on compact domains, i.e.,
\begin{equation}
\left| p_{\text{NQS}}(\phi_i) - p^*(\phi_i) \right| < \epsilon, \quad \forall \phi_i \in \{0,1\}^n, \text{ for any } \epsilon > 0.
\end{equation}
This makes NQS a powerful surrogate for quantum states, particularly when access to real quantum measurement probabilities is limited, costly, or corrupted by finite-shot noise.
However, NQS have no intrinsic access to quantum entanglement or nonlocal correlations unless such structure is explicitly learned from data or specified. 

In this work, we reinterpret NQS as the classical backbone in our HPQS framework. The NQS serves as a flexible estimator that complements the partial and noisy output of PQCs. This neural component enables the model to infer unmeasured amplitudes or denoise quantum observations, thereby reducing the burden on quantum resources while retaining model expressivity.

\section{Hybrid Parameterized Quantum States}

\subsection{General Formulation}

To improve the robustness and expressivity of variational learning under finite-shot or noisy measurement conditions, we propose a hybrid framework called \textbf{Hybrid Parameterized Quantum States (HPQS)}. This framework blends quantum-generated information with classical neural estimators based on NQS. 

Let \( \hat{p}_i^{\text{emp}} \) denote the empirical estimate of \( |\langle \phi_i | \psi(\boldsymbol{\theta}) \rangle|^2 \) obtained from finite-shot quantum measurements of PQC. We define postprocessed quantum and classical predictions as:
\begin{equation}
\Tilde{p}_i^{(Q)} = G(\phi_i, \hat{p}_i^{\text{emp}}), \quad \Tilde{p}_i^{(C)} = H(\phi_i, f_{\boldsymbol{\gamma}}(\phi_i)),
\end{equation}
where \( G: (\phi_i, \hat{p}_i^{\text{emp}}) \mapsto \mathbb{R} \) is a postprocessing function applied to quantum outputs (e.g., MLP decoder or tensor contraction), \( H: (\phi_i, f_{\boldsymbol{\gamma}}(\phi_i)) \mapsto \mathbb{R} \) is a postprocessing function applied to classical NQS outputs (e.g., alignment layer or task-adaptive transformation).
 
The final hybrid prediction for basis state \( \phi_i \) is defined as:
\begin{equation}
\hat{p}_i = \lambda \cdot \Tilde{p}_i^{(Q)} + (1 - \lambda) \cdot \Tilde{p}_i^{(C)},
\end{equation}
with \( \lambda \in [0, 1] \) as a scalar blending coefficient, which is considered as a hyperparameter. 
This unified formulation captures a wide range of hybrid strategies, such as:
\begin{itemize}
    \item Direct blending when \( G(x) = x \) and \( H(x) = x \).
    \item Hybrid decoding pipelines when \( G \) and \( H \) are expressive decoders.
    \item When \( \lambda = 1 \), the model reduces to a pure quantum (PQC-based) formulation using only empirical measurement.
    \item When \( \lambda = 0 \), the model reduces to a purely classical (NQS-based) estimation framework.
\end{itemize}
With more possibilities investigated in Appendix~\ref{app:otherquantum}. By introducing \( G \) and \( H \), HPQS enables the quantum and classical branches to be flexibly adapted to task-specific modalities or scales. It allows their outputs to be composed in a representation space before blending. A schematic drawing of HPQS is shown in Fig.~\ref{fig:scheme}. 

From the perspective of the quantum circuit, this blended formulation allows PQC-derived outputs to be regularized and enhanced through classical estimation, mitigating finite-shot uncertainty and measurement noise. From the perspective of the classical neural estimator, HPQS introduces physically grounded quantum information into the learning process, enriching the model’s expressivity with entanglement and non-classical correlations. Together, this hybrid mechanism enables scalable and noise-resilient variational learning, well suited for NISQ-era quantum systems, as demonstrated in the empirical experiment section.

Since HPQS includes an NQS component with universal approximation capacity, it inherits the ability to approximate any target probability distribution \( p^*(\phi_i) \) over computational basis states. Specifically, for any \( \epsilon > 0 \), there exist parameters \( \boldsymbol{\phi}_{GH}\) and a blending weight \( \lambda \in [0, 1] \) such that the hybrid estimate satisfies

\begin{equation}
|\hat{p}_i - p^*(\phi_i)| < \epsilon, \quad \forall \phi_i \in \{0,1\}^n,
\end{equation}
where \( \hat{p}_i = \lambda \cdot \Tilde{p}_i^{(Q)} + (1 - \lambda) \cdot \Tilde{p}_i^{(C)} \) is the HPQS prediction, \( \boldsymbol{\phi}_{GH}\) are the corresponding parameters of $\hat{p}_i$ (constructing $G$, $H$, $f_{\gamma}$, and $\hat{p}_i^{\text{emp}}$). This theoretical expressivity complements the robustness brought by this parameterization design, especially under finite-shot or noisy conditions.

\subsection{Parameter Generation via HPQS in Quantum-Train}

In the QT framework \cite{liu2024quantum, liu2024training, liu2024qtrl, lin2024quantum}, the goal is to generate the full set of trainable parameters \( \boldsymbol{a} = (a_1, a_2, \dots, a_m) \) of a classical neural network using a hybrid quantum-classical architecture. This replaces the need to train these parameters directly and instead formulates parameter generation as a variational learning task over a lower-dimensional latent space.
Let \( m \) be the total number of parameters in the target neural network. A PQC with \( n = \lceil \log_2 m \rceil \) qubits and \( L \) layers is constructed to define the quantum state 
\begin{equation}
\label{eq:ansatz}
|\psi(\boldsymbol{\theta}) \rangle = \left(\prod_{i=1}^{n-1} \text{CNOT}^{i, i+1} \prod_{j=1}^{n} R_Y^j (\theta_{j}^{(L)}) \right)^L |0\rangle^{\otimes n}.
\end{equation}
Here, each single-qubit rotation gate $R_Y^j$ is parameterized by $\theta_j^{(L)}$, where $j$ indexes the qubits, and $L$ represents the circuit depth. The controlled-NOT ($\text{CNOT}$) gates create entanglement between qubits.
After measurement, we obtain a finite-shot empirical estimate \( \hat{p}_i^{\text{emp}} \) of the probability for each computational basis state \( \phi_i \in \{0, 1\}^n \). At the same time, an NQS \( f_{\boldsymbol{\gamma}}(\phi_i) \) is trained as a classical neural estimator for the outcomes of the same basis set.
To map basis states and their associated quantum or classical probabilities to parameter values \( a_i \in \boldsymbol{a} \), both branches are passed through postprocessing functions based on a tensor network architecture:
\begin{equation}
\Tilde{a}_i^{(Q)} = G_{\text{TN}}(\phi_i, \hat{p}_i^{\text{emp}}), \quad \Tilde{a}_i^{(C)} = H_{\text{TN}}(\phi_i, f_{\boldsymbol{\gamma}}(\phi_i)).
\end{equation}
\( G_{\text{TN}} \) and \( H_{\text{TN}} \) are tensor network-based mapping models \cite{liu2024quantum2}, applied separately to the quantum and classical branches (see Appendix~\ref{app:hp} for architectural details). Both functions can be viewed as structured decoders that learn to map basis-probability pairs to real-valued neural parameters.

The final HPQS-generated parameter is given by:
\begin{equation}
a_i = \lambda \cdot \Tilde{a}_i^{(Q)} + (1 - \lambda) \cdot \Tilde{a}_i^{(C)},
\end{equation}
with a scalar blending coefficient \( \lambda \in [0, 1] \). This combination allows the model to leverage noisy but informative quantum measurements while compensating with a classical neural estimator where quantum information is limited.
Once all \( a_i \) are generated via HPQS, they are used to instantiate the full set of weights in a classical neural network, which is then evaluated on a downstream task such as image classification. Only the quantum parameters \( \boldsymbol{\theta} \), classical parameters \( \boldsymbol{\phi} \), and tensor network weights in \( G_{\text{TN}} \) and \( H_{\text{TN}} \) are optimized. The number of trainable parameters is thus polylogarithmic in \( m \) \cite{liu2024quantum}, making this approach highly parameter-efficient.
The use of tensor networks as the mapping model ensures expressive yet scalable transformation from basis information to model parameters. This design also allows seamless batching and parallelism during training, and unifies the interpretation of classical and quantum-generated states.

\subsection{Expectation-Based Quantum Machine Learning with HPQS}

In conventional QML, variational quantum circuits are often trained to minimize a loss function based on the expectation value of an observable \cite{schuld2021effect}. For classification tasks, this typically involves encoding input data into a quantum state, applying a PQC, and measuring observables to obtain scalar values (e.g., logits or class scores). The predicted model output is then derived from these expectation values, following the approach in prior work \cite{wang2022quantumnas}.

We extend this approach through the HPQS framework by blending quantum-derived expectation values with classical predictions from a neural network model. In this setup, the quantum branch processes input \( x \) by preparing a parameterized quantum state \( \ket{\psi(\boldsymbol{\theta}, x)} \) with angle encoding method, followed by measurement of a Pauli observable \( \hat{Z} \) to yield the empirical expectation value:
\begin{equation}
\hat{y}_{\text{quantum}} = \hat{\mathbb{E}}[\hat{Z}] = \sum_i z_i \hat{p}_i^{\text{emp}}, \quad \hat{p}_i^{\text{emp}} = \text{empirical estimate of } |\langle \phi_i | \psi(\boldsymbol{\theta}, x) \rangle|^2.
\end{equation}
This scalar output may be noisy or imprecise under finite-shot conditions, especially for deep circuits or large input spaces.
Simultaneously, a classical neural network \( f_{\boldsymbol{\gamma}}(x) \) is trained to produce a prediction from the same input \( x \). Rather than choosing one branch exclusively, HPQS blends the two via postprocessing functions:
\begin{equation}
\Tilde{y}^{(Q)} = G(\hat{y}_{\text{quantum}}), \quad \Tilde{y}^{(C)} = H(f_{\boldsymbol{\gamma}}(x)),
\end{equation}
where \( G \) is a postprocessing function applied to the quantum expectation value (e.g., scaling or nonlinearity), \( H \) is a transformation of the classical output (e.g., projection, batch normalization, or logit shaping). The actual $H$ and $G$ used in this example are described in the Appendix~\ref{app:hp}.

The final HPQS prediction is given by:
\begin{equation}
\hat{y} = \lambda \cdot \Tilde{y}^{(Q)} + (1 - \lambda) \cdot \Tilde{y}^{(C)},
\end{equation}
where \( \lambda \in [0, 1] \) is a blending coefficient. This structure allows HPQS to operate as a hybrid inference model, using quantum expectation values when reliable and falling back on classical learning when quantum resources are sparse or noisy.
The output \( \hat{y} \) is used as input to a loss function (e.g., cross-entropy for classification or mean squared error for regression), and gradients are propagated through both the quantum and classical branches. This hybrid formulation unifies PQC-based learning with classical machine learning in a scalable and shot-efficient manner.

This configuration is particularly well-suited for low-qubit, expectation-based quantum learning tasks such as binary classification. As demonstrated in Section~\ref{sec:experiments}, HPQS achieves more stable and improved learning performance compared to PQC-only baselines under finite-shot conditions, while remaining competitive with purely classical methods. Notably, HPQS mitigates the shot inefficiency commonly observed in PQC by directly processing task inputs, reflecting its core principle of seamlessly blending quantum measurements with classical inference.

\subsection{Enhancing LLM Fine-Tuning with HPQS-based Quantum Parameter Adaptation}

Large language models (LLMs) such as GPT-2 and Gemma-2 \cite{radford2019language, team2024gemma}  have demonstrated strong generalization across a wide range of natural language tasks. However, fine-tuning such models remains computationally expensive due to their large number of trainable parameters. To address this, parameter-efficient fine-tuning (PEFT) methods, such as Low-Rank Adaptation (LoRA) \cite{hu2022lora}, adapt only a small subset of parameters while freezing the rest of the model.
Quantum Parameter Adaptation (QPA) \cite{liu2024quantum3} extends the QT approach to the PEFT setting, to generate only the adaptation-specific parameters. Unlike full model generation, QPA focuses on fine-tuning modules such as LoRA matrices, which are significantly smaller in dimension but crucial for downstream task adaptation.
In QPA, the LoRA update to a pretrained weight matrix \( W_0 \in \mathbb{R}^{d \times k} \) is expressed as:
\begin{equation}
W_0 + \Delta W = W_0 + B A,
\end{equation}
where \( A \in \mathbb{R}^{r \times k} \), \( B \in \mathbb{R}^{d \times r} \), and \( r \ll \min(d, k) \) defines the rank of the low-rank update. The goal of QPA is to generate the entries of \( A \) and \( B \) via a hybrid quantum-classical procedure, while minimizing the number of trainable variables in the quantum circuit and classical neural estimator.
To scale QPA to large models, we adopt a batched parameter generation strategy \cite{liu2024federated} as in the original QPA proposal. The full adaptation vector \( \boldsymbol{a} \in \mathbb{R}^{r(d + k)} \) is partitioned into \( n_{\text{ch}} \) chunks of size \( n_{\text{mlp}} \), with each chunk generated independently using HPQS. For each chunk index \( i \), a basis $\phi_i$ for a quantum state $|\psi(\boldsymbol{\theta}) \rangle$ is prepared and measured to yield \( \hat{p}_i^{\text{emp}} \), while an NQS estimator provides the corresponding classical value \( f_{\boldsymbol{\gamma}}(\phi_i) \). These values are then passed through tensor network decoders \( G \) and \( H \) as in the QT setup:
\begin{equation}
\label{eq:tn1}
\Tilde{a}_i^{(Q)} = G_{\text{TN}}(\phi_i, \hat{p}_i^{\text{emp}}), \quad \Tilde{a}_i^{(C)} = H_{\text{TN}}(\phi_i, f_{\boldsymbol{\gamma}}(\phi_i)),
\end{equation}
and the HPQS estimate of the adaptation parameters becomes:
\begin{eqnarray}
&&\boldsymbol{a} = (\hat{a}_1, \hat{a}_2, \hdots, \hat{a}_{n_{\text{ch}}}), \\
&&\hat{a}_i = \lambda \cdot \Tilde{a}_i^{(Q)} + (1 - \lambda) \cdot \Tilde{a}_i^{(C)}, \quad \forall i \in \{ 1, 2, \ldots, n_{\text{ch}} \}\\
&& \hat{a}_i = ( a_{i,1}, a_{i,2}, \dots, a_{i,j} ), \quad \forall j \in \{ 1, 2, \ldots, n_{\text{mlp}} \}.
\end{eqnarray}
This process is repeated across all \( n_{\text{ch}} \) chunks to produce the complete set of LoRA weights. Importantly, this approach reduces the number of required qubits from $N = \lceil \log_2 m \rceil$ to
\begin{equation}
\label{eq:QTBG_qubits_usage}
N = \lceil \log_2 n_{\text{ch}} \rceil = \lceil \log_2 \left(  \lceil \frac{m}{n_{\text{mlp}}}\rceil \right) \rceil,
\end{equation} allowing adaptation to large-scale models even under tight quantum hardware constraints. Additionally, the shot efficiency and noise robustness of HPQS enable stable fine-tuning with limited quantum samples, addressing a key limitation of the original QPA formulation, as noted in Appendix G of that work.

\section{Empirical Experiments}
\label{sec:experiments}

\subsection{Overall Performance}
To validate the effectiveness of the proposed HPQS framework, we conduct experiments across three representative quantum machine learning tasks~\footnote{The code is attached to the supplemental material.}: parameter generation (QT), expectation-based classification (QML), and parameter-efficient adaptation (QPA). For each task, we compare four configurations:
\begin{itemize}
\item \textbf{PQC (exact)}: full access to the quantum state vector via exact simulation.
\item \textbf{PQC (finite)}: probabilities estimated from a finite number of measurement shots simulation.
\item \textbf{NQS}: pure classical neural quantum state.
\item \textbf{HPQS (finite)}: hybrid design combining NQS and PQC (finite).
\end{itemize}

Each experiment is conducted under both noise-free and IBM quantum noise models (see Appendix~\ref{app:noise}), using a fixed shot budget for PQC-based configurations (e.g., \(10 \times 2^n\), where \(2^n\) corresponds to the Hilbert space size, denoted as HSS). All simulations and optimizations are implemented using PyTorch and TorchQuantum \cite{wang2022quantumnas, paszke2017automatic}.

\textbf{Parameter Generation via HPQS in Quantum-Train.} We evaluate HPQS in the context of full parameter generation for classical neural networks. The quantum output is postprocessed through a tensor-network-based mapping model and blended with its NQS counterpart. The task involves classifying the MNIST dataset \cite{lecun2010mnist} (10 classes) using a compact convolutional neural network (CNN) with 6690 target parameters, which are generated from a significantly smaller set of trainable variables. As shown in Table~\ref{tab:qt_model_comparison}, HPQS achieves accuracy comparable to PQC (exact), while significantly outperforming PQC (finite) and approaching or exceeding the performance of NQS under limited-shot conditions. Each reported accuracy is averaged over three random seeds, and the detailed architecture and training settings are provided in Appendix~\ref{app:hp}.

\begin{table}[h]
\centering
\caption{Comparison of QT models on MNIST-10 classification under various access and shot regimes.}
\begin{tabular}{|l|c|c|l|}
\hline
\multicolumn{4}{|c|}{\textbf{QT (MNIST-10)}} \\
\hline
\textbf{Model} & \textbf{Testing Accuracy (\%)} & \textbf{\# Training Param.} & \textbf{Shot Count} \\
\hline
PQC (exact)     & \textbf{90.59 $\pm$ 1.12} & 904 & $\infty$ \\
HPQS (finite)   & 86.28 $\pm$ 0.58 & 649 & 10 $\times$ HSS \\
HPQS (finite)   & \textbf{87.64 $\pm$ 0.47}$^\dagger$ & 792 & 10 $\times$ HSS \\
NQS            & 76.13 $\pm$ 7.25 & 512 & — \\
NQS            & 84.80 $\pm$ 5.45 & 2624 & — \\
PQC (finite)    & 57.54 $\pm$ 5.61 & 1164 & 10 $\times$ HSS \\
PQC (finite)    & 65.83 $\pm$ 2.56 & 1424 & 10 $\times$ HSS \\
\hline
\end{tabular}
\\
\label{tab:qt_model_comparison}
\end{table}

\textbf{Expectation-Based Quantum Machine Learning with HPQS.} We evaluate HPQS on a binary classification task involving MNIST digits 3 versus 6, following the setup described in \cite{wang2022quantumnas}. In this setting, a variational quantum classifier generates scalar outputs based on measured expectation values. HPQS blends these quantum-derived values with classical predictions from a neural estimator. As shown in Table~\ref{tab:qml_model_comparison}, HPQS outperforms both NQS and PQC under finite-shot conditions, achieving a peak accuracy of \( \mathbf{90.66 \pm 1.09\%} \) using 20$\times$HSS shots. Notably, it surpasses the performance of PQC (exact), which achieves \( 86.66 \pm 2.17\% \), and significantly outperforms PQC (finite), which degrades substantially under limited-shot settings (e.g., \( 46.21 \pm 9.26\% \) with 5$\times$ HSS). Each result is averaged over three random seeds. These findings highlight HPQS’s ability to enhance robustness against shot noise and deliver superior performance compared to either component individually under practical quantum constraints.

\textbf{Enhancing LLM Fine-Tuning with HPQS-based Quantum Parameter Adaptation.} To demonstrate the scalability of HPQS, we apply it to QPA for LoRA-based fine-tuning of large language models, including GPT-2 (80M) and Gemma-2 (2B), on the WikiText-2 dataset \cite{merity2016pointer}. The low-rank matrices used for adaptation are generated via a batched HPQS parameter generation scheme. Fine-tuning is applied only to the final linear projection layer, with the remainder of the model kept frozen, as in the original QPA paper \cite{liu2024quantum3}.

As shown in Table~\ref{tab:qpa_model_comparison}, HPQS achieves strong performance under limited quantum resources. On GPT-2, HPQS (finite) obtains a perplexity of 4.612 using only 1$\times$ HSS shots, compared to 1.616 for PQC (exact), 6.975 for PQC (finite), and a much worse 153.629 for NQS. Similarly, on Gemma-2, HPQS (finite) achieves 1.467, closely matching PQC (exact) at 1.427, while outperforming both NQS (1.471) and PQC (finite) (1.472).
Each reported number reflects the best result over three random seeds. These results confirm that HPQS can maintain competitive or superior performance compared to pure quantum or classical baselines, even when constrained to finite-shot quantum access. This demonstrates HPQS’s effectiveness as a practical, scalable approach for low-resource fine-tuning in modern LLMs.

\subsection{Robustness to Quantum Noise}
We further evaluate the noise resilience of HPQS by simulating real IBM quantum noise models. Across all tasks, HPQS consistently outperforms PQC (finite) under noisy conditions, with relative degradation that is notably less severe than pure PQC models. This confirms HPQS’s robustness in practical, imperfect quantum environments. Detailed comparisons are provided in Appendix~\ref{app:noise}.

\begin{table}[h]
\centering
\caption{Comparison of QML models under different access and shot regimes.}
\begin{tabular}{|l|c|c|l|}
\hline
\multicolumn{4}{|c|}{\textbf{QML (MNIST-2)}} \\
\hline
\textbf{Model} & \textbf{Testing Accuracy (\%)} & \textbf{\# Training Param.} & \textbf{Shot Count} \\
\hline
PQC (exact)     & 86.66 $\pm$ 2.17 & 40 & $\infty$ \\
HPQS (finite)   & 88.97 $\pm$ 5.08 & 65 & 5 $\times$ HSS \\
HPQS (finite)   & \textbf{90.66 $\pm$ 1.09} & 65 & 20 $\times$ HSS \\
NQS             & \textbf{89.67 $\pm$ 5.30}$^\dagger$ & 21 & — \\
PQC (finite)    & 46.21 $\pm$ 9.26 & 40 & 5 $\times$ HSS \\
PQC (finite)    & 53.33 $\pm$ 6.61 & 40 & 20 $\times$ HSS \\
\hline
\end{tabular}
\\
\label{tab:qml_model_comparison}
\end{table}

\begin{table}[h]
\centering
\small 
\caption{Comparison of QPA models under different access and shot regimes.}
\begin{tabular}{|l|l|c|c|l|}
\hline
\textbf{Task} & \textbf{Model} & \textbf{Testing PPL } & \textbf{\# Training Param.} & \textbf{Shot Count} \\
\hline
QPA (GPT-2 \& Wikitext-2) & PQC (exact)     & \textbf{1.616} & 22368 & $\infty$    \\
QPA (GPT-2 \& Wikitext-2) & HPQS (finite)     & \textbf{4.612}$^\dagger$ & 23585 & 1 $\times$ HSS    \\
QPA (GPT-2 \& Wikitext-2) & NQS              & 153.629 & 18113 & —                   \\
QPA (GPT-2 \& Wikitext-2) & PQC (finite)      & 6.975 & 22368 & 1 $\times$ HSS   \\
\hline
QPA (Gemma-2 \& Wikitext-2) & PQC (exact)     & \textbf{1.427} & 140432 & $\infty$    \\
QPA (Gemma-2 \& Wikitext-2) & HPQS (finite)     & \textbf{1.467}$^\dagger$ & 141585 & 1 $\times$ HSS    \\
QPA (Gemma-2 \& Wikitext-2) & NQS              & 1.471 & 136321 & —                   \\

QPA (Gemma-2 \& Wikitext-2) & PQC (finite)      & 1.472 & 140432  & 1 $\times$ HSS   \\
\hline
\end{tabular}
\label{tab:qpa_model_comparison}
\end{table}

\section{Discussion and Conclusion}

In this work, we introduced \textit{Hybrid Parameterized Quantum States (HPQS)}, a unified framework that interpolates between PQC and NQS for variational quantum learning. By blending quantum measurement outcomes with classical neural estimators, HPQS enables scalable, shot-efficient, and noise-resilient learning suitable for near-term quantum devices. We validate HPQS across three representative tasks spanning increasing model complexity: expectation-based classification in QML, full parameter generation in QT, and fine-tuning of large language models in QPA. In all cases, we empirically demonstrate that HPQS outperforms PQC under finite-shot conditions and achieves comparable or better performance than PQC (exact) and NQS, while maintaining strong expressivity and robustness to statistical and hardware-induced noise.

\paragraph{Limitations and Future Work.}
While HPQS mitigates many challenges of variational quantum learning, its performance depends on the effectiveness of the classical estimator and the quality of quantum measurements. In low-shot regimes, improper blending or weak estimation can limit improvement over classical baselines. Future directions include learning task-adaptive blending mechanisms, incorporating uncertainty-aware estimators, and extending HPQS to more expressive quantum encodings and structured datasets such as molecular graphs. Furthermore, HPQS can be integrated with quantum error mitigation techniques or classical generative priors, to further enhance performance.

Our work suggests a principled and practical solution for bridging classical and quantum representations in variational quantum learning. It provides a foundation for future hybrid architectures that adaptively combine quantum resources with classical models, without requiring fully quantum inference, which is essential for enabling real-world applications on both noisy intermediate-scale and fault tolerant quantum hardware.

\clearpage

\bibliographystyle{unsrt}
\bibliography{nips.bib, bib/qst.bib, bib/qml.bib, bib/qem.bib, bib/nqs.bib, bib/qt, bib/llm}

\clearpage
\appendix

\section{Effects of Quantum Computer Noise}
\label{app:noise}

\begin{table}[h]
\centering
\caption{Comparison of QML models under different noise settings. (shot count $=20 \times$  HSS)}
\begin{tabular}{|l|c|c|l|}
\hline
\multicolumn{4}{|c|}{\textbf{QML (MNIST-2)}} \\
\hline
\textbf{Model} & \textbf{Testing Accuracy (\%)} & \textbf{\# Training Param.} & \textbf{Noise Model}  \\
\hline
HPQS (finite)   & 88.97 $\pm$ 5.08 & 65 & — \\
HPQS (finite)   & \textbf{93.33 $\pm$ 1.08} & 65 & ibm\_fez  \\
HPQS (finite)   & \textbf{90.21 $\pm$ 3.50}$^{\dagger}$ & 65 & ibm\_torino  \\
PQC (finite)    & 53.33 $\pm$ 6.61 & 40 &  —   \\
PQC (finite)    & 44.44 $\pm$ 4.12 & 40 & ibm\_fez \\
PQC (finite)    & 55.10 $\pm$ 6.65 & 40 & ibm\_torino  \\
\hline
\end{tabular}
\\

\label{tab:qml_model_comparison_noisy}

\end{table}

\begin{table}[h]
\centering
\caption{Comparison of QT models under different noise settings. (shot count $=10 \times$  HSS)}
\begin{tabular}{|l|c|c|l|}
\hline
\multicolumn{4}{|c|}{\textbf{QT (MNIST-10)}} \\
\hline
\textbf{Model} & \textbf{Testing Accuracy (\%)} & \textbf{\# Training Param.} & \textbf{Noise Model} \\
\hline
HPQS (finite)   & 86.28 $\pm$ 0.58 & 649 & —  \\
HPQS (finite)   & 77.04 $\pm$ 7.79 & 649 & ibm\_fez  \\
HPQS (finite)   & 79.58 $\pm$ 4.88 & 649 & ibm\_torino  \\

HPQS (finite)   & \textbf{87.64 $\pm$ 0.47} & 792 & — \\
HPQS (finite)   & \textbf{87.04 $\pm$ 0.52}$^{\dagger}$ & 792 & ibm\_fez \\
HPQS (finite)   & 81.62 $\pm$ 4.58 & 792 & ibm\_torino \\

PQC (finite)    & 57.54 $\pm$ 5.61 & 1164 & — \\
PQC (finite)    & 61.94 $\pm$ 5.49 & 1164 & ibm\_fez \\
PQC (finite)    & 56.11 $\pm$ 2.87 & 1164 & ibm\_torino  \\

PQC (finite)    & 65.83 $\pm$ 2.56 & 1424 & —  \\
PQC (finite)    & 55.89 $\pm$ 4.39 & 1424 & ibm\_fez  \\
PQC (finite)    & 56.82 $\pm$ 4.07 & 1424 & ibm\_torino \\
\hline
\end{tabular}
\\
\label{tab:qt_model_comparison_noise_noisy}
\end{table}

\begin{table}[h]
\centering
\caption{Comparison of QPA models (GPT-2) under different noise settings. (shot count $=1 \times$  HSS)}

\begin{tabular}{|l|c|c|l|}
\hline
\multicolumn{4}{|c|}{\textbf{QPA (GPT-2 \& Wikitext-2)}} \\
\hline
\textbf{Model} & \textbf{Testing PPL } & \textbf{\# Training Param.} & \textbf{Noise Model} \\
\hline
HPQS (finite)     & \textbf{4.612} & 23585 &   —  \\
HPQS (finite)     & 53.565 & 23585 &  ibm\_fez   \\
HPQS (finite)     & 53.785 & 23585 &  ibm\_torino   \\

PQC (finite)      & \textbf{6.975}$^{\dagger}$ & 22368 &  —  \\
PQC (finite)      & 86.695 & 22368 &  ibm\_fez   \\
PQC (finite)      & 88.369 & 22368 & ibm\_torino  \\
\hline
\end{tabular}
\label{tab:qpa_model_comparison_noisy}
\end{table}

\begin{table}[h]
\centering
\caption{Comparison of QPA models (Gemma-2) under different noise settings. (shot count $=1 \times$  HSS)}

\begin{tabular}{|l|c|c|l|}
\hline
\multicolumn{4}{|c|}{\textbf{QPA (Gemma-2 \& Wikitext-2)}} \\
\hline
\textbf{Model} & \textbf{Testing PPL } & \textbf{\# Training Param.} & \textbf{Noise Model} \\
\hline
HPQS (finite)     & \textbf{1.467} & 141585 &   —  \\
HPQS (finite)     & \textbf{1.470}$^{\dagger}$ & 141585 &  ibm\_fez   \\
HPQS (finite)     & 1.472 & 141585 &  ibm\_torino   \\

PQC (finite)      & 1.472 & 140432 &  —  \\
PQC (finite)      & 1.472 & 140432 &  ibm\_fez   \\
PQC (finite)      & 1.472 & 140432 & ibm\_torino  \\
\hline
\end{tabular}
\label{tab:qpa_model_comparison_noise_gemma_2}
\end{table}

To assess the noise robustness of HPQS, we evaluate its performance under realistic quantum noise simulation using IBM’s available noise models~\footnote{IBM Quantum provides noise models based on the properties of real hardware backends:\url{https://docs.quantum.ibm.com/api/qiskit/0.19/qiskit.providers.aer.noise.NoiseModel}.} {ibm\_torino} and {ibm\_fez}. Across three representative settings: expectation-based QML, QT, and QPA. Each task is executed with a fixed shot budget, specifically $20\times$ HSS for QML, $10\times$ HSS for QT, and $1\times$ HSS for QPA.

\paragraph{Expectation-based QML HPQS under Noise setting.}
As shown in Table~\ref{tab:qml_model_comparison_noisy}, HPQS maintains high classification accuracy under both noise-free and noisy conditions. For instance, it achieves 93.33\% accuracy under ibm\_fez and 90.21\% under ibm\_torino, both of which surpass the corresponding PQC (finite) results that degrade to 44.44\% and 55.10\%, respectively. Notably, HPQS also outperforms the PQC (exact) baseline (86.66\%) in the noise-free case.

\paragraph{QT with HPQS under Noise setting.}
In the QT scenario (Table~\ref{tab:qt_model_comparison_noise_noisy}), HPQS achieves 87.64\% testing accuracy without noise and retains strong performance under noisy settings (87.04\% with ibm\_fez and 81.62\% with ibm\_torino). In contrast, PQC (finite) experiences significant degradation, dropping to accuracies between 55\% and 65\% depending on noise level and parameter count.

\paragraph{QPA with HPQS under Noise setting.}
For QPA (Table~\ref{tab:qpa_model_comparison_noisy} and Table~\ref{tab:qpa_model_comparison_noise_gemma_2}), which targets LoRA-based fine-tuning of large language models, HPQS consistently achieves lower perplexity than PQC (finite) under both noise-free and noisy conditions. In the GPT-2 experiment, HPQS yields 4.612 in the noise-free setting, while the noisy variants (ibm\_fez and ibm\_torino) reach 53.565 and 53.785, respectively. These values remain substantially better than PQC (finite), which suffers from notable degradation, reaching 86.695 to 88.369 under noise. A similar trend is observed in the Gemma-2 experiment, where HPQS achieves 1.467 to 1.472 perplexity across all conditions, marginally outperforming PQC (finite).

\section{Implementation Details and Training Hyperparameter Configuration}
\label{app:hp}
In this section, we provide the training hyperparameter configuration used for the results presented in the main text. All experiments were conducted on a single NVIDIA V100S GPU with 32GB VRAM.

\subsection{Tensor Network Mapping via Matrix Product States}

Tensor networks (TNs) have emerged as powerful tools for modeling high-dimensional correlations in quantum many-body physics, classical machine learning, and quantum circuit simulations. Their compact and structured representation makes them especially well-suited for capturing the relationship between quantum measurement distributions and classical neural network parameters.

Following the supervised learning formulation of matrix product states (MPS) in \cite{stoudenmire2016supervised}, we parameterize the mapping function from quantum-classical inputs to model weights using a tensor decomposition. Specifically, the output vector $\vec{\omega} \in \mathbb{R}^d$ is computed via a contraction between an MPS weight tensor W and an input-dependent feature map $\Xi(\mathbf{x})$, where:
\begin{equation}
W_{s_1, s_2, \ldots, s_{N+1}} = \sum_{\boldsymbol{\alpha}} A^{\alpha_1}{s_1} A^{\alpha_1 \alpha_2}{s_2} \cdots A^{\alpha_N}{s_{N+1}},
\end{equation}
and $\mathbf{x} = (x_1, x_2, \ldots, x_{N+1}) \in \mathbb{R}^{N+1}$ represents the concatenated input composed of the computational basis state $|\phi\rangle$ and the associated measurement probability $|\langle \phi | \psi(\boldsymbol{\theta}) \rangle|^2$. The corresponding feature map is expressed as:
\begin{equation}
\Xi^{s_1, s_2, \ldots, s_{N+1}}(\mathbf{x}) = \xi^{s_1}(x_1) \otimes \xi^{s_2}(x_2) \otimes \cdots \otimes \xi^{s_{N+1}}(x_{N+1}),
\end{equation}
where each local map $\xi^{s_j}(x_j) \in \mathbb{R}^2$ is defined by:
\begin{equation}
\xi^{s_j}(x_j) =
\begin{bmatrix}
x_j \\
1 - x_j
\end{bmatrix}.
\end{equation}
The final output of the tensor network is obtained as:
\begin{equation}
\label{eq:tn2}
G(\mathbf{x}) = W \cdot \Xi(\mathbf{x}) = \vec{\omega}.
\end{equation}
Here, the MPS core tensors $A^{\alpha_j}_{s_j}$ are the learnable parameters of the model, with bond dimension $r$ governing the expressivity. This approach allows for structured generalization and efficient parameter scaling in the mapping of hybrid quantum-classical features to downstream model components.

Returning to Eq.~\ref{eq:tn1} in the main text, the tensor network mapping functions applied to the quantum and classical branches are expressed as:
\begin{equation}
\Tilde{a}_i^{(Q)} = G_{\text{TN}}(\phi_i, \hat{p}_i^{\text{emp}}), \quad \Tilde{a}_i^{(C)} = H_{\text{TN}}(\phi_i, f_{\boldsymbol{\gamma}}(\phi_i)).
\end{equation}

These tensor network functions are structurally aligned with the contraction form presented in Eq.~\ref{eq:tn2}, where the resulting output vector \( \vec{\omega} \) corresponds to either \( \Tilde{a}_i^{(Q)} \) or \( \Tilde{a}_i^{(C)} \), depending on whether it originates from the quantum or classical branch. In our implementation, the bond dimensions are task-specific: for the Quantum-Train experiments, the quantum-side tensor network \( G_{\text{TN}} \) is assigned a bond dimension of \( r = 2 \), while the classical-side \( H_{\text{TN}} \) adopts a minimal structure with \( r = 1 \). In contrast, for the QPA experiments involving large-scale language models, we employ a more expressive configuration with \( r = 10 \) for \( G_{\text{TN}} \) and \( r = 4 \) for \( H_{\text{TN}} \). These mapping modules are implemented using the \texttt{TorchMPS} library \cite{torchmps}, which provides efficient support for matrix product state (MPS) computations in PyTorch.

\subsection{Parameter Settings for Expectation-Based QML Experiments}

We evaluate HPQS in an expectation-based QML task, closely following the setup in \cite{wang2022quantumnas}. The quantum branch utilizes a 4-qubit PQC, corresponding to a HSS of $2^4 = 16.$ Input MNIST images are downsampled to $4 \times 4$ using PyTorch’s avg\_pool2d operation. The resulting 16-dimensional vectors are mapped to gate angles using a sequence of rotation gates: 4RY, 4RZ, 4RX, and 4RY.
Measurements are performed in the Pauli-Z basis, producing expectation values in $[-1, 1]$ for each qubit. For binary classification (digits 3 vs 6), the expectation values of qubits $\{0,1\}$ and $\{2,3\}$ are summed independently to form two logits, which are then normalized using a softmax function to yield class probabilities (corresponding to postprocessing $G$).

In the classical branch (NQS), the same $4 \times 4$ image is reshaped into a 16-dimensional vector and passed through a fully connected layer of shape $(16, 1)$, followed by a ReLU activation and a final linear layer of shape $(1, 2)$ to produce logits. The quantum and classical outputs are blended using a fixed coefficient $\lambda = 0.1$.
The entire hybrid model is trained using the negative log-likelihood (NLL) loss over 5 epochs, optimized with Adam at a learning rate of $5 \times 10^{-3}$. Each reported result is averaged over three random seeds.

\subsection{Parameter Settings for Quantum-Train Experiments}

For the QT task, we adopt a setup aligned with the original framework proposed by \cite{liu2024quantum}. The objective is to generate the full parameter set of a compact CNN comprising 6690 trainable parameters. To match this capacity, the quantum branch uses \( n = \lceil \log_2 6690 \rceil = 13 \) qubits, resulting in a HSS of \( 2^{13} = 8192 \).
Unlike the expectation-based QML setting, the classification task here spans all 10 MNIST digit classes. Full \( 28 \times 28 \) grayscale images are used as input to the CNN. The quantum state is prepared using a layered parameterized ansatz defined as:
\begin{equation}
\label{eq:ansatz}
|\psi(\boldsymbol{\theta}) \rangle = \left(\prod_{i=1}^{n-1} \text{CNOT}^{i, i+1} \prod_{j=1}^{n} R_Y^j (\theta_{j}^{(L)}) \right)^L |0\rangle^{\otimes n},
\end{equation}
where \( L=1 \) denotes the number of layers, each composed of entangling CNOT gates and single-qubit rotations.

The classical branch (NQS) processes the binary representation of each basis state \( \phi_i \in \{0,1\}^n \) through a fully connected layer of shape \( (n, 32) \), followed by a ReLU activation and a final linear layer of shape \( (32, 1) \), yielding a scalar probability estimate. The quantum and classical branches are blended using a fixed coefficient \( \lambda = 0.5 \).
Training is performed for 50 epochs using the Adam optimizer with a learning rate of \( 1 \times 10^{-4} \). The loss function is categorical cross-entropy, as appropriate for multi-class classification. All results are averaged over three random seeds for statistical reliability.

\subsection{Parameter Settings for Quantum Parameter Adaptation}

For the QPA experiments, we largely follow the setup proposed in the original QPA work~\cite{liu2024quantum3}, as summarized in Table~\ref{tab:qpa_hp}. In this setting, \( \alpha \) denotes the scaling factor used in the low-rank adaptation (LoRA) scheme. The chunk size for batched parameter generation is set to \( n_{\text{mlp}} = 512 \) for GPT-2 and \( n_{\text{mlp}} = 4096 \) for Gemma-2. Given this chunking strategy, the required number of qubits is determined by (Eq.~\ref{eq:QTBG_qubits_usage}):
\begin{eqnarray}
&&n_{\text{GPT-2}} = \left\lceil \log_2 \left\lceil \frac{204{}100}{512} \right\rceil \right\rceil = 9, \\
&&n_{\text{Gemma-2}} = \left\lceil \log_2 \left\lceil \frac{1{}032{}192}{4096} \right\rceil \right\rceil = 8,
\end{eqnarray}

where 204{}100 and 1{}032{}192 are the number of trainable parameters in the LoRA module (\( r = 4 \)) for the final linear layers of GPT-2 and Gemma-2, respectively.
The quantum ansatz employed in this setting mirrors the layered structure used in the QT experiments, with a circuit depth of \( L = 8 \). The classical NQS branch adopts the same architecture as in the QT setup, consisting of a fully connected network that operates on the binary representation of basis states. The outputs of the quantum and classical branches are combined using a fixed blending coefficient of \( \lambda = 0.5 \).

\begin{table}[htbp]
    \centering
    \caption{Hyperparameter configurations of QPA LoRA for fine-tuning GPT-2 and Gemma-2 with WikiText-2 dataset.}
    \vspace{10pt} 
    \small
    \begin{tabular}{lcc}
        \toprule
        \multirow{2}{*}{\textbf{Hyperparameters}} & \multicolumn{2}{c}{\textbf{QPA LoRA}} \\
         & \textbf{GPT-2} & \textbf{Gemma-2 } \\
        \midrule
        LoRA Rank $r$             &   \multicolumn{2}{c}{4} \\
        $\alpha$             &     \multicolumn{2}{c}{2r}   \\  
        Dropout              & 0.05   & 0.0  \\
        Optimizer            & \multicolumn{2}{c}{AdamW} \\
        LR                   & \multicolumn{2}{c}{1e-5} \\
        LR Scheduler         & \multicolumn{2}{c}{Linear} \\
        Batch size           & \multicolumn{2}{c}{1}  \\
        Warmup Steps         & \multicolumn{2}{c}{0}  \\
        Epochs               & \multicolumn{2}{c}{3}\\
        \bottomrule
    \end{tabular}

    \label{tab:qpa_hp}

\end{table}

\section{Related Quantum Learning Frameworks as Special Cases of HPQS}
\label{app:otherquantum}

The HPQS framework defines a hybrid variational quantum model by blending postprocessed outputs from a quantum circuit and a classical neural estimator:
\begin{equation}
\hat{p}_i = \lambda \cdot G(\phi_i, \hat{p}_i^{\text{emp}}) + (1 - \lambda) \cdot H(\phi_i, f_{\boldsymbol{\gamma}}(\phi_i)),
\end{equation}
where:
\begin{itemize}
\item $\hat{p}_i^{\text{emp}}$ is the empirical probability estimate from quantum measurements
\item $f_{\boldsymbol{\gamma}}(\phi_i)$ is the NQS-predicted value
\item $G(\cdot)$ and $H(\cdot)$ are postprocessing functions (can be both quantum or classical)
\item $\lambda \in [0, 1]$ controls the contribution of each branch
\end{itemize}

This formulation naturally subsumes a range of quantum learning models:

\paragraph{PQC-Only Learning and NQS-Only Learning.} As discussed extensively in the main text, HPQS recovers conventional variational quantum models in the limiting case of \( \lambda = 1 \) with a simple postprocessing function \( G \). This corresponds to purely PQC-based learning, such as in variational quantum algorithms (VQAs) \cite{cerezo2021variational}, where optimization relies entirely on quantum circuit evaluations. Conversely, when \( \lambda = 0 \) and simple \( H \), HPQS reduces to a NQS formulation \cite{carleo2017solving}. These limiting cases highlight the generality of HPQS as a unifying framework for both purely quantum and purely classical learning paradigms.

\paragraph{Hybrid Quantum-Classical Learning.} Several models in QML apply classical neural layers directly on top of quantum features. For instance, Quanvolutional Neural Networks \cite{henderson2020quanvolutional} use quantum circuits as feature extractors followed by classical classifiers. These can be seen as cases where $G(\cdot)$ includes a learnable classical transformation and $\lambda = 1$. 

The framework introduced in \cite{mari2020transfer} appends PQCs after classical neural network layers, effectively enabling transfer learning from classical to quantum domains. Within the HPQS formulation, this can be interpreted as a special case where the blending coefficient is set to \( \lambda = 0 \), and the postprocessing function in the classical branch \( H(\cdot) \) includes a learnable quantum circuit applied to the classical output. This highlights how quantum layers can act as task-adaptive decoders within the broader HPQS architecture.

\paragraph{Quantum-Classical Embeddings.} Recent work on quantum-enhanced feature spaces, such as Quantum Support Vector Machines \cite{havlivcek2019supervised}, implicitly postprocess quantum features via classical similarity or metric functions. These frameworks correspond to HPQS with fixed or partially learnable G and $\lambda = 1$.

\paragraph{Classical Postprocessing and Noise Mitigation.} Classical estimators are increasingly used to compensate for quantum noise. Notable examples include error mitigation via neural networks and other approaches \cite{strikis2021learning, kim2023evidence}. These methods naturally fit into the HPQS architecture as specific instantiations of $G(\cdot)$, and highlight the flexibility of HPQS in NISQ settings.

HPQS provides a general and expressive framework that unifies diverse quantum-classical learning paradigms. Its modular structure supports flexible combinations of classical and quantum components, with the blending coefficient \( \lambda \) offering explicit control over their respective contributions. The incorporation of postprocessing functions \( G \) and \( H \) further enhances the adaptability of each branch to task-specific requirements. While the current formulation considers a fixed \( \lambda \) across the entire system, future extensions may explore \textit{context-aware} \( \lambda_i \) values, allowing the model to dynamically adjust the quantum-classical trade-off per input or per output component. This offers a promising direction for developing more adaptive and data-efficient hybrid architectures.

\section{Why Increasing Qubit Count Leads to Higher Estimation Error}
\label{app:err}
We extend the analysis in the main text to formally characterize how measurement uncertainty scales with the number of qubits. Let \( \hat{P}(|\phi_i\rangle) \) denote the empirical estimate of the probability of observing basis state \( |\phi_i\rangle \), based on \( n'_{\text{shot}} \) measurement repetitions. Here, \( n'_{\text{shot}} = n_{\text{shot}} / 2^n \) represents the average number of shots allocated per basis state in an \( n \)-qubit system with Hilbert space size \( 2^n \).

Hoeffding’s inequality provides a probabilistic bound on the deviation of this empirical estimate from its true expectation:
\begin{equation}
P\left( \left| \hat{P}(|\phi_i\rangle) - \mathbb{E}[P(|\phi_i\rangle)] \right| \geq \epsilon \right) \leq 2 \exp(-2 \epsilon^2 n'_{\text{shot}}).
\end{equation}

Rewriting the bound in terms of a fixed confidence level \( \delta \in (0, 1) \), we obtain:
\begin{equation}
\epsilon = \sqrt{ \frac{1}{2 n'_{\text{shot}}} \ln \left( \frac{2}{\delta} \right) } = \sqrt{ \frac{2^n}{2 n_{\text{shot}}} \ln \left( \frac{2}{\delta} \right) }.
\end{equation}

This reveals a key limitation: under a fixed total shot budget \( n_{\text{shot}} \), the estimation error \( \epsilon \) grows exponentially with the number of qubits \( n \). In other words, as the Hilbert space expands, fewer measurements are allocated per basis state, leading to larger statistical fluctuations. This scaling behavior underscores the importance of shot-efficient designs, such as HPQS, which integrate classical estimators to reduce reliance on dense quantum measurements.

\clearpage

\end{document}